\documentclass[doublecol]{epl2} 
\usepackage[T1]{fontenc}
\usepackage{mathptmx}
\DeclareMathAlphabet{\mathcal}{OMS}{cmsy}{m}{n} 
\usepackage{multirow}
\usepackage[font=small]{caption}
\usepackage{float}
\usepackage{mathrsfs}
\usepackage{color}
\usepackage{graphicx}
\usepackage{subeqnarray}
\usepackage{dcolumn}
\usepackage{bm}

\title{Routes to stratified turbulence and temporal intermittency \\ revealed by a cluster-based network model \\ of experimental data}

\shorttitle{Stratified turbulence and intermittency: a cluster-based network model} 

\author{Adrien Lefauve\inst{1}\!\!\footnote{\,Email:  lefauve@damtp.cam.ac.uk } \and Yui Hin Marvil Cheung\inst{1} \and Xianyang Jiang\inst{1} \and Miles M. P. Couchman\inst{2}}
\shortauthor{Lefauve, Cheung, Jiang \& Couchman}

\institute{                    
  \inst{1}Department of Applied Mathematics and Theoretical Physics, University of Cambridge,  Cambridge, CB3 0WA, UK\\
  \inst{2}Department of Mathematics and Statistics, York University, Toronto, ON M3J 1P3, Canada
}

\abstract{  Modelling fluid turbulence using a `skeleton' of coherent structures has traditionally progressed by focusing on a few canonical laboratory experiments such as pipe flow and Taylor-Couette flow. We here consider the stratified inclined duct, a sustained shear flow whose density stratification allows for the exploration of a wealth of new coherent and intermittent states at significantly higher Reynolds numbers than in unstratified flows. We automatically identify the underlying turbulent skeleton of this experiment with a  data-driven method combining dimensionality reduction and unsupervised clustering of shadowgraph visualisations. We demonstrate the existence of multiple types of turbulence across parameter space and intermittent cycling between them, revealing distinct transition pathways. With a cluster-based network model of intermittency we uncover patterns in the transition probabilities and residence times under increasing levels of turbulent dissipation. Our method and results pave the way for new reduced-order models of multi-physics turbulence.}

\begin{document}

\maketitle

\textbf{Introduction.} --- One of the greatest challenges in fluid dynamics is identifying reduced-order descriptions of the high-dimensional, strongly nonlinear dynamical system describing turbulence. One promising approach, coherent-structure modeling, dates back to \cite{hopf_mathematical_1948,ruelle_nature_1971} and assumes that the turbulent dynamics of practical importance are low-dimensional, with phase-space trajectories spending significant time near a  small set, or `skeleton', of exact (though usually unstable) solutions of the Navier-Stokes equations, called simple invariant solutions or exact coherent states \cite{kawahara_significance_2012,graham_exact_2021}. This description enables the prediction of turbulent statistics using a weighted average over these solutions \cite{cvitanovic_recurrent_2013}, and can reproduce some aspects of the spatiotemporal coherence of the fully-nonlinear dynamics \cite{lucas_irreversible_2017,page_revealing_2021}. 

The search for coherent turbulent skeletons requires a deep understanding of the transition 
from laminar to turbulent flow. The study of hydrodynamic stability dates back to O. Reynolds \cite{reynolds_experimental_1883}, who introduced in 1883 an experiment that has since become a dominant paradigm: cylindrical pipe flow. A key insight was that the transition to turbulence was governed by the Reynolds number $\textrm{Re}=uh/\nu$, where $u$ and $h$ denote the flow's characteristic velocity and length scales and $\nu$ denotes the fluid's kinematic viscosity. Recent reviews \cite{eckhardt_turbulence_2007,willis_experimental_2008,barkley_theoretical_2016,avila_transition_2023} highlight the significant progress made on understanding the route to turbulence and the skeleton of pipe flow over the last 150 years. 
A crucial step in this journey was made by G. I. Taylor \cite{taylor_stability_1923}, who introduced in 1923 a second canonical experiment: the flow between two concentric rotating cylinders, now known as Taylor-Couette flow.  
Over the past century,  pipe and Taylor-Couette flows revealed two fundamentally different routes to turbulence with increasing $\textrm{Re}$. Pipe flow is linearly stable, and nonlinearities amplify finite-amplitude disturbances (a subcritical transition) into turbulent puffs and slugs of increasing lifetime like in excitable and bistable media \cite{barkley_theoretical_2016}. By contrast, Taylor-Couette flow (when dominated by inner-cylinder rotation) is linearly unstable, and nonlinearities lead to the saturation of exponential 
instabilities (a supercritical transition) and turbulence after a sequence of successive instabilities \cite{coles_transition_1965,feldman_routes_2023}. 

\begin{figure}[t]
\includegraphics[width=0.9\linewidth]{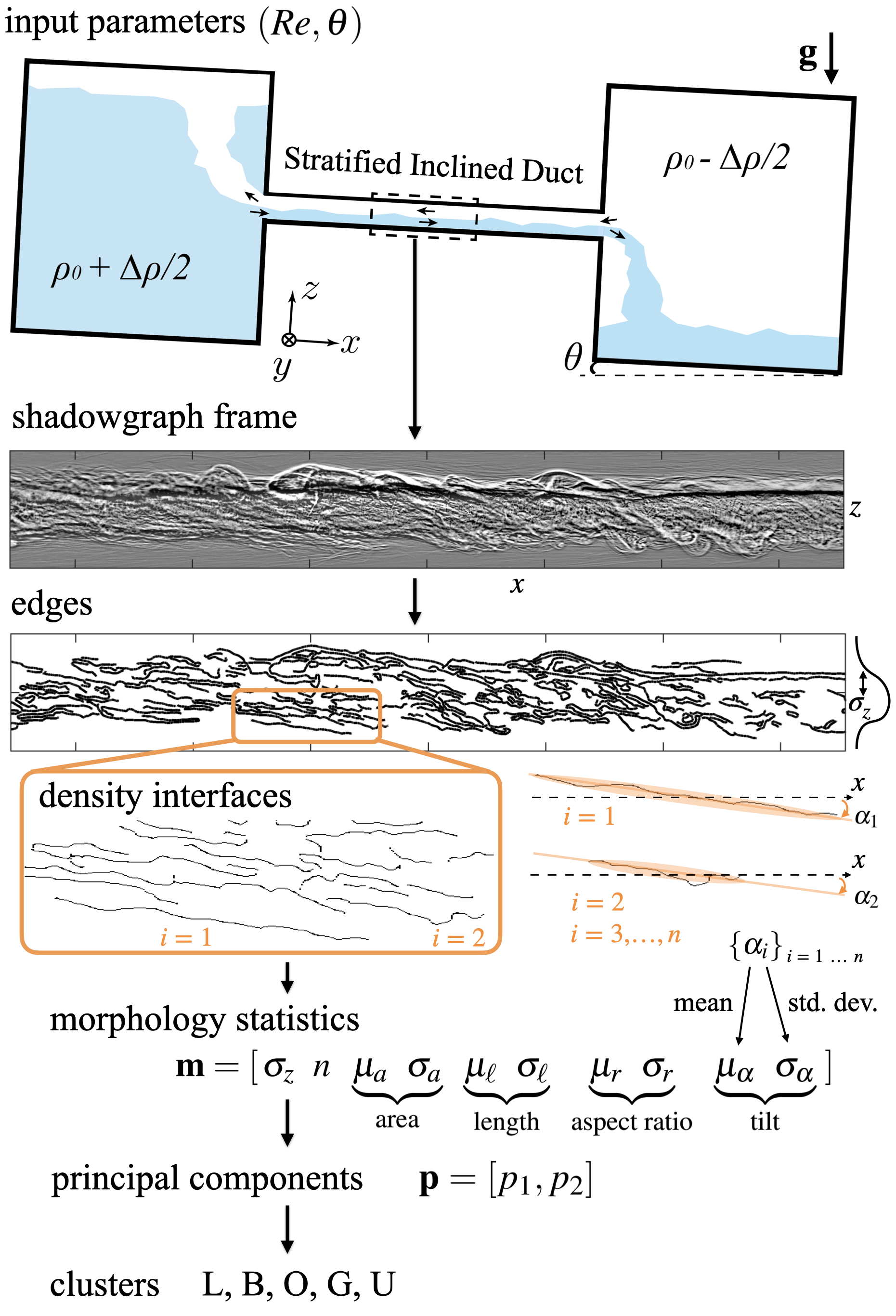}
\caption{Experimental setup (the duct is 2 m long, 100 mm wide and $H=50$ mm tall), shadowgraph data and  dimensionality reduction pipeline: edge detection, 10-dimensional (10-D) morphology vector, 2-D vector in principal coordinates, and unsupervised clustering  \cite{lefauve_data-driven_2024}.}
\label{fig:pipeline}
\end{figure}

We here consider routes to turbulence in a third and comparatively less-well-known laboratory experiment: the `stratified inclined duct' (SID, see top of fig.~\ref{fig:pipeline}). This density-stratified flow features rich transitional and intermittent dynamics which, we will argue, provide a new fruitful paradigm for advancing turbulence modelling. Previous progress in characterising turbulence has crucially relied upon the study of transitional flows exhibiting spatio-temporal intermittency, which is notoriously difficult to grasp. In SID, stratification introduces stabilising effects and thus a second dimensionless parameter, yielding richer intermittent behaviors at higher $\textrm{Re}$ \cite{turner_buoyancy_1973,deusebio_intermittency_2015}  
and thus more generic building blocks for the skeleton of turbulence \cite{lucas_layer_2017,salehipour_deep_2019,smith_turbulence_2021}. 

The addition of two-layer stratification to the shear flow in a tilted pipe was first done by Reynolds in his 1883 paper \cite[\S\,12]{reynolds_experimental_1883}, yielding a turbulent transition that was fundamentally different to pipe flow. The SID experiment was born later \cite{macagno_interfacial_1961,meyer_stratified_2014}, by connecting a long rectangular duct to two large saltwater reservoirs of different densities  $\rho_0 \pm \Delta\rho/2$ (fig.~\ref{fig:pipeline}). SID has two control parameters: $\textrm{Re}$ based on half the duct height $h=H/2$ and the layer-averaged buoyancy velocity scale $u=\sqrt{gH\Delta\rho/\rho_0}$, and the tilt angle $\theta$, which provides extra energy \cite{lefauve_regime_2019} to sustain dissipative states for longer periods and explore their spatio-temporal intermittency. A rich `intermittent regime' was characterized by \cite{meyer_stratified_2014,lefauve_regime_2019,lefauve_buoyancy_2020} in intermediate regions of parameter space $(\textrm{Re},  \theta)$, between a finite-amplitude Holmboe wave regime at low values of $\textrm{Re} \, \theta$ (weakly turbulent, with little interfacial mixing) and a fully turbulent regime at high $\textrm{Re} \, \theta$ (which never relaminarises and has intense mixing). Novel experiments measuring the time-resolved, three-dimensional velocity and density fields \cite{partridge_versatile_2019} allowed for the development of a basic `SID skeleton', demonstrating that (i) the transition to turbulence was supercritical, mediated by the `confined' Holmboe instability \cite{lefauve_structure_2018}, and (ii) three-dimensional coherent structures in the fully turbulent regime  (e.g. hairpin vortices) could be traced back to this linear instability \cite{jiang_evolution_2022}. However,  intermittency has long remained enigmatic.

\textbf{Outline.} -- In this Letter, we employ a novel data-driven approach for advancing the discovery and understanding of the fundamental states and intermittent dynamics underpinning SID turbulence across $(\textrm{Re},\theta)$ space. Such an analysis previously relied upon a trained human eye identifying 
qualitative flow features within experimental videos \cite{meyer_stratified_2014,lefauve_buoyancy_2020}, limiting the accuracy, repeatability, and feasibility of classifying large datasets, and hence its success to date. These limitations have recently been overcome \cite{lefauve_data-driven_2024} with an objective method, combining dimensionality reduction and unsupervised clustering, for classifying flow snapshots based on embedded dynamical structures. We review and emphasise how this method identifies a variety of distinct turbulent states distinguished by their finescale density stratification. 
We reveal unique routes to turbulence by studying the distribution of these states in $(\textrm{Re},\theta)$ and the intermittent temporal cycles between them. A key contribution of this Letter is our subsequent focus on this 
intermittency. By developing a cluster-based network model of these reduced-order  dynamics, we reveal new statistical patterns in cluster transitions and residence times between transitions.

\textbf{Dataset and dimensionality reduction.} --- Our dataset consists of 50,155 individual shadowgraph frames (see example in fig.~\ref{fig:pipeline}) belonging to 113 movies \cite{jiang_research_2023_EPL}. Each movie visualises the evolution of a sustained, sheared, salt-stratified turbulent flow in SID  for a fixed $(\textrm{Re},\theta)$ over hundreds of advective (or shear) time units $h/u$.
Collectively, the movies span the Holmboe wave, intermittent and fully turbulent human-classified regimes \cite{lefauve_buoyancy_2020} across a wide region of parameter space $\textrm{Re}=300-5000$, $\theta=1-6^\circ$.  
The greyscale intensity of the shadowgraphs is approximately proportional to the $x-z$ curvature of the fluid's refractive index, and hence the density field, integrated over the spanwise direction $y$ of the light rays. The dataset is thus well-suited for studying the structure and temporal evolution of density interfaces embedded within the flow, which are known to be energetically and dynamically meaningful \cite{linden_mixing_1979, caulfield_layering_2021,couchman_mixing_2023}. %
The automated dimensionality reduction pipeline recently introduced by \cite{lefauve_data-driven_2024} (sketched in fig.~\ref{fig:pipeline}) begins by detecting density interfaces in shadowgraph frames using a Canny edge-detection algorithm \cite{canny_computational_1986}. Only high-contrast edges are detected, to ensure that experimental noise does not influence our results. The properties of each connected density interface are then computed: number per frame $n$, lists of respective areas $\{a_i\}_{i=1,\ldots,n}$,  lengths $\{\ell_i\}$, aspect ratios $\{r_i\}$ and tilt angles $\{\alpha_i\}$ based on the fitting of an ellipse (orange, fig. 1). A 10-dimensional morphology vector
\begin{equation}\label{eq:m}
    \mathbf{m}_f=[ \sigma_z \ \  n \ \ \mu_a \ \ \sigma_a \ \ \mu_\ell \ \ \sigma_\ell \ \ \mu_r \ \ \sigma_r \ \ \mu_\alpha \ \ \sigma_\alpha]
\end{equation}
is then constructed, representing each frame $f$ by its number of interfaces $n$, the vertical standard deviation of edge pixel density $\sigma_z$, and the mean ($\mu$) and standard deviation ($\sigma$) of its four lists of interface properties. 
This low-dimensional representation remains physically-interpretable based on the structure of density filaments embedded within the flow, unlike other compression algorithms such as auto-encoders where the reduced-order ``latent space'' may be harder to interpret. A principal component analysis (PCA) \cite{brunton_data-driven_2019} is used to identify correlations within the matrix $\mathbf{M}=[\mathbf{m}_f]_{f=1,\ldots,50155}$, allowing the dataset to be further reduced to a two-dimensional (2-D) PCA basis, explaining $\approx 80$\,\% of the total variance. Our pipeline thus compresses each 1.5-MPixel shadowgraph frame into a 2-D vector $\mathbf{p}_f$ in PCA space $(P_1,P_2)$. More details can be found in \cite{lefauve_data-driven_2024}.

\begin{figure}[t]
\includegraphics[width=0.87\linewidth]{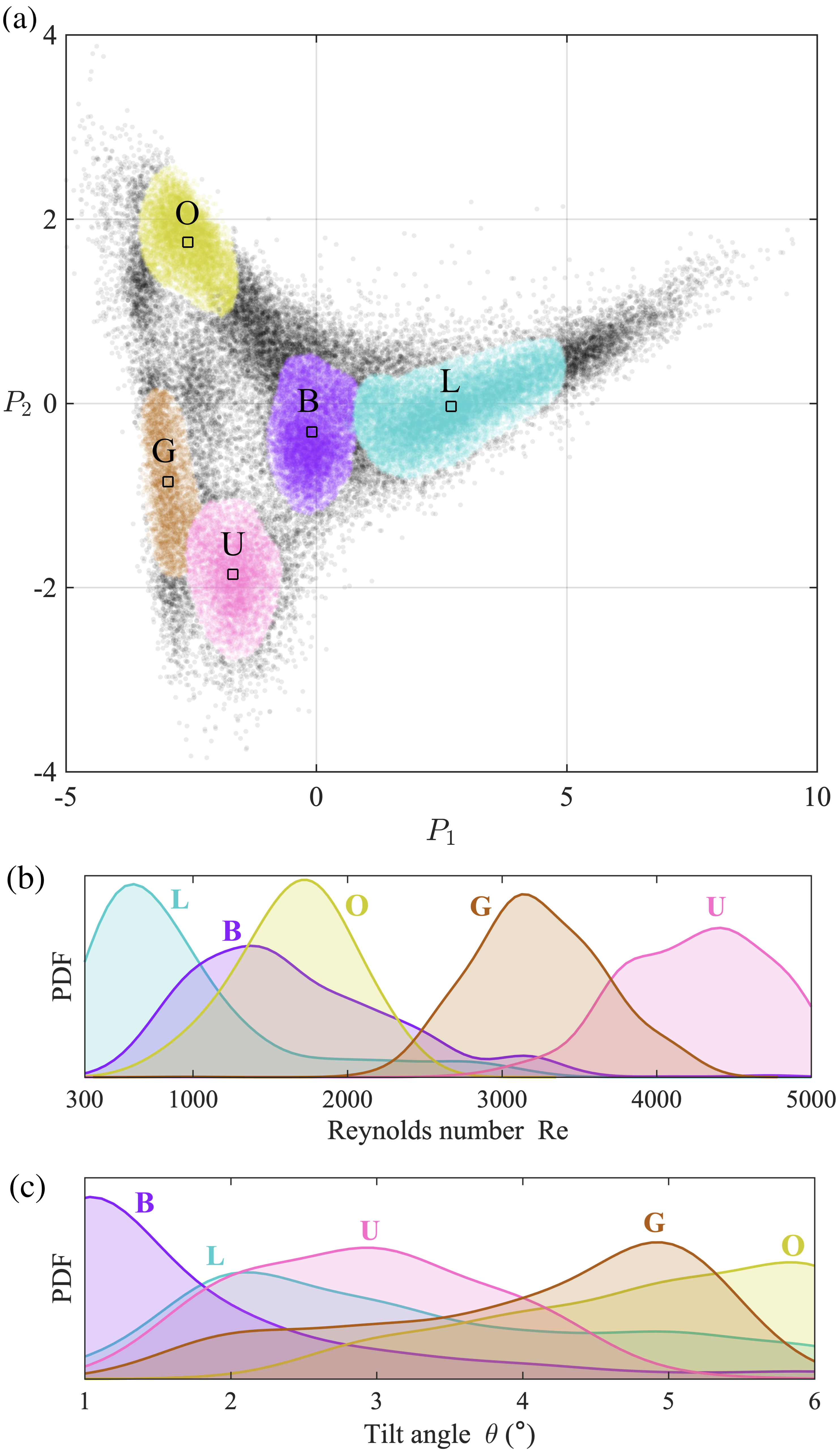}
\caption{Clustered data:  All 50,155 experimental frames in the space of principal components (a). Points belonging to clusters are coloured, while sparser, unclustered points are in black (adapted from \cite{lefauve_data-driven_2024}). Smoothed probability distributions of clusters in $\textrm{Re}$ (b) and $\theta$ (c). }
\label{fig:clusters}
\end{figure}

\textbf{Classification and physical interpretation.} -- The  density-based clustering algorithm `OPTICS' \cite{ankerst_ordering_1999,couchman_data-driven_2021} is then used to automatically detect dense clusters in the PCA space $\mathbf{P}$ (fig.~\ref{fig:clusters}a). Five clusters are detected, accounting for $\approx 80$~\% of the data, with unclustered points indicating sparser regions. OPTICS, based on the seminal algorithm DBSCAN, was chosen for its simplicity and distinct advantages over other algorithms: it automatically determines the number of clusters, recognises clusters of arbitrary shape and density, and is robust to noise \cite{han_data_2011}. Figures~\ref{fig:clusters}b-c show the distributions of $\textrm{Re}$ and $\theta$ for the frames belonging to each cluster. The successive prevalence of each cluster L, B, O, G and U as $\textrm{Re}$ increases, and of a different succession of clusters as $\theta$ increases, already suggests that the classification captures meaningful and non-trivial physics. 

\begin{figure}[t]
\includegraphics[width=0.848\linewidth]{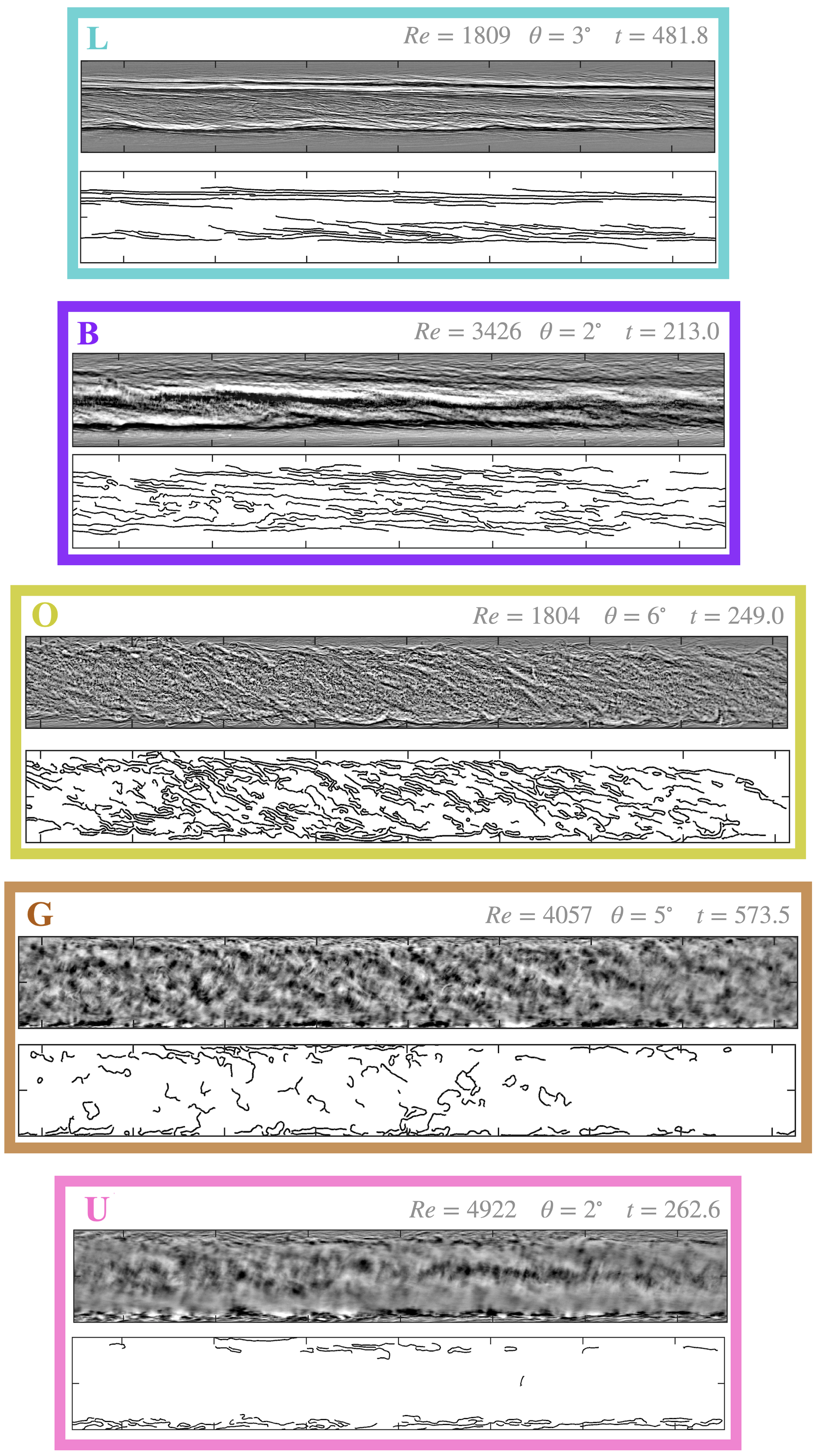}
\caption{Frames corresponding to the centroid of each cluster (squares in fig.~\ref{fig:clusters}a) representing the five types of turbulence. Shadowgraphs are shown (top) with the detected edges (bottom). Note the dimensionless $\textrm{Re},\theta$ and snapshot time $t$ (top right). Adapted from \cite{lefauve_data-driven_2024}.}
\label{fig:examples}
\end{figure}

The clusters are interpreted in terms of their physical flow properties in fig.~\ref{fig:examples}, using a representative frame from each cluster's centroid. We summarise these properties here to set the stage for our further analysis, noting that more detail can be found in \cite{lefauve_data-driven_2024}.
Frame L intuitively qualifies as the most laminar state (for reference, a hypothetical single laminar interface would yield principal coordinates far off the top right vertex in fig.~\ref{fig:clusters}a). It represents what we may call \textit{laminarising turbulence}, where small-scale structure from a preceding turbulent phase remains visible in the shadowgraph snapshot, but the pattern of edges shows a `stacking' of flat and stable density interfaces. By contrast, frame B illustrates \textit{braided turbulence}, owing to its pair of central `braids' with relatively two-dimensional curvature in the density field evidenced by a strong contrast in shadowgraph intensity, representing two strong interfaces surrounded by a number of weaker ones. Frame O illustrates \textit{overturning turbulence}, owing to its numerous short, tilted and hence unstable interfaces. Frames G and U illustrate \textit{granular} and \textit{unstructured turbulence} respectively, where small-scale three-dimensional turbulent motions result in blurred images due to averaging in the cross-duct $y$ direction.
The similarity of clusters G and U is consistent with them being adjacent in fig.~\ref{fig:clusters}a;  the more detailed OPTICS output in \cite{lefauve_data-driven_2024} demonstrates that they are sub-components of a larger unifying cluster, but are distinguishable based on subtle differences in their density interfaces. Specifically, the turbulence in G is more granular, owing to a slightly stronger contrast and hence more detected edges, especially at mid-height, whereas the turbulence is less structured in U, with flatter edges localised near the top and bottom walls where the blurred mixing layer meets the more quiescent boundary layers.

\begin{figure}[t!]
\includegraphics[width=\linewidth]{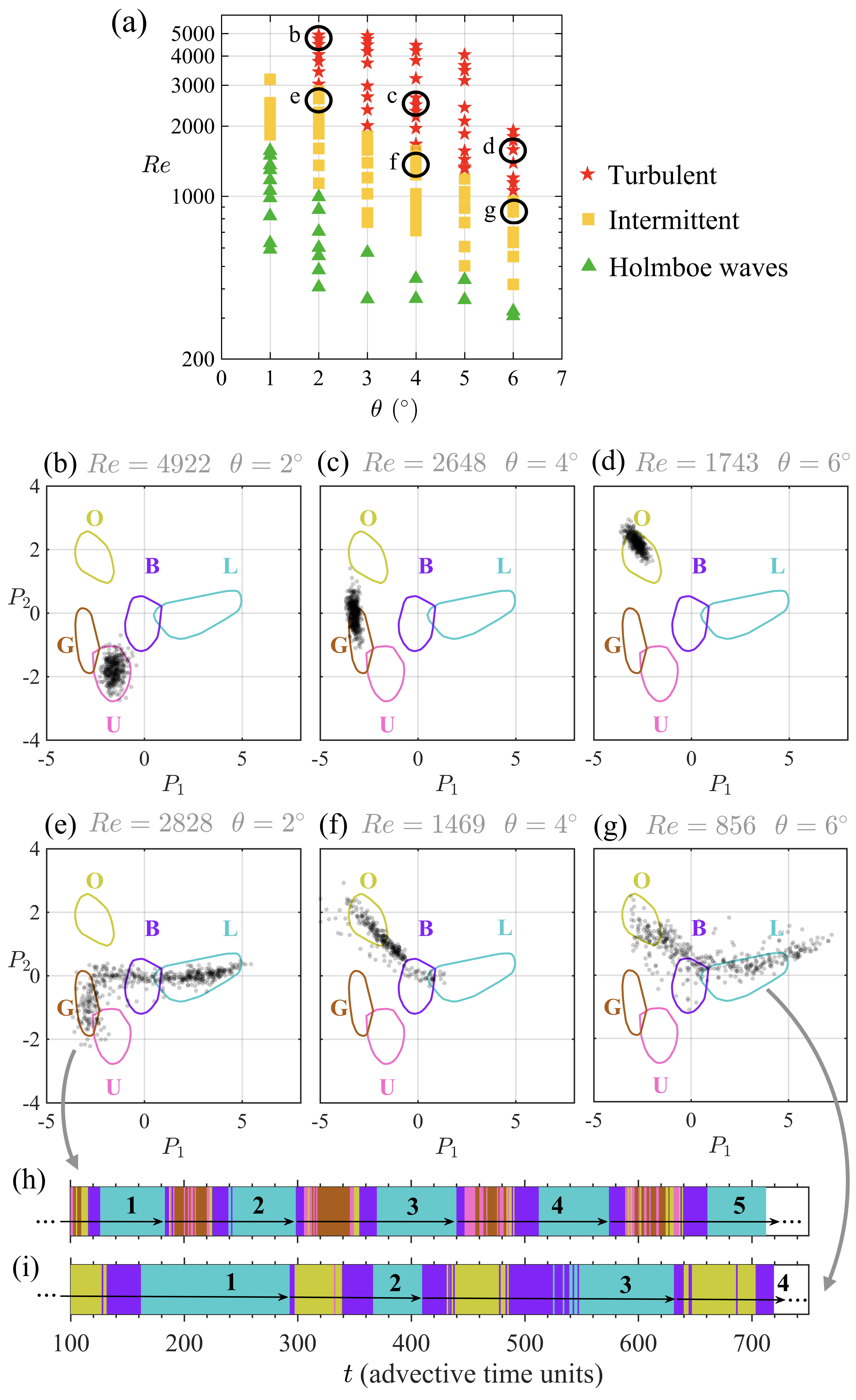}
\caption{Temporal dynamics (b-i) in six experiments chosen from the  113 experiments in the space of input parameters (a). The frames are shown in (b-g) by translucent symbols and the cluster boundaries are in colours. The cycling time series of (e,g) are shown in (h,i) respectively (without initial transients $t<100$), with cycle numbers in bold. Unclustered frames are coloured based on the nearest cluster.}
\label{fig:timeseries}
\end{figure}

\textbf{Phase-space dynamics and temporal intermittency.} --- We now highlight in fig.~\ref{fig:timeseries} the phase space trajectories $(P_1,P_2)$ of six experiments, spanning a range of $(\textrm{Re},\theta)$ (fig.~\ref{fig:timeseries}a). We compare three experiments traditionally classified (by the human eye) as fully `Turbulent' at $\textrm{Re} \, \theta \approx 10,000$ (fig.~\ref{fig:timeseries}b-d) and three classified as `Intermittent' at $\textrm{Re} \, \theta \approx 6000$ (fig.~\ref{fig:timeseries}e-g). We note that the product $\textrm{Re} \, \theta$ is proportional to  the time- and volume-averaged rate of turbulent kinetic energy dissipation  $\epsilon$ in the flow \cite{lefauve_regime_2019}, and to the dynamic range of stratified turbulence, or buoyancy Reynolds number $\textrm{Re}_b=(L_O/L_K)^{4/3}$, measuring the separation between the Ozmidov lengthscale $L_O=(\epsilon/N^3)^{1/2}$ and  the Kolmogorov lengthscale $L_K=(\nu^3/\epsilon)^{1/4}$  ($N$ is the averaged buoyancy frequency) \cite[Sec. 5.1]{lefauve_experimental2_2022}.

Our analysis reveals that the most dissipative turbulence, traditionally classified as fully developed turbulence  (top $\textrm{Re}\,\theta$ line of highest dynamic range), is in fact characterised by different states. At high $\textrm{Re}$ and low $\theta=2^\circ$ (fig.~\ref{fig:timeseries}b) turbulence is exclusively unstructured, staying within or very near cluster U. At intermediate $\textrm{Re}$ and $\theta=4^\circ$ (fig.~\ref{fig:timeseries}c), it shifts to being granular in cluster G with excursions to the sparser (unclustered) space towards O. At low $\textrm{Re}$ and high $\theta=6^\circ$ (fig.~\ref{fig:timeseries}d), it is exclusively of overturning type, in or near cluster O. 
By contrast, less dissipative turbulence along the bottom $\textrm{Re}\,\theta$ line transitions between clusters over time (fig.~\ref{fig:timeseries}e-i). At intermediate $\textrm{Re}$ and $\theta$ (fig.~\ref{fig:timeseries}f), the turbulence is of mixed braided/overturning type, confined to clusters B, O and the intervening space, with rare relaminarisations. At high $\textrm{Re}$ and low $\theta$ (fig.~\ref{fig:timeseries}e,h), the route to turbulence is fundamentally different: the time series is now quasi-periodic across L, B and G. The cycles, numbered in bold, have periods $T \approx 120-140$ advective (shear) time units. Each cycle starts with a very short excursion  to braided turbulence (residence time $\approx 5$), followed by a long visit in or near granular turbulence ($\approx 50$), then another excursion  to braided turbulence ($\approx 10$), eventually leading to a long relaminarisation ($\approx 65$). At low $\textrm{Re}$ and high $\theta$ (fig.\ref{fig:timeseries}g,i), the route to turbulence is different again, being quasi-periodic across L, B and O (with a more variable period $T\approx 120-220$). Importantly, the intense turbulent periods are of overturning (O) rather than granular (G) type. However they are again accessed by the same braided (B) `gateway', albeit by `turning up' via its top end of rather by `turning down' via its middle (compare fig.~\ref{fig:timeseries}g and e). Moreover, in both cases the excursions through B during the relaminarisation phases G/O$\rightarrow$B$\rightarrow$L are consistently longer than during the unstable transitional phases L$\rightarrow$B$\rightarrow$G/O. These results suggest that the dynamics of SID intermittency organise around two inherently different `slow manifolds', motivating the future exploration of a higher-dimensional phase space to resolve the bursting and relaxation dynamics in the orthogonal `fast manifold' \cite[\S~6.7.2]{schmid_data-driven_2021}. These cycles are reminiscent of the lifecycle of a (transient) 
Kelvin-Helmholtz billow \cite{mashayek_goldilocks_2021,smith_turbulence_2021}, but SID flow evidently harbours a greater wealth of turbulent attractors and transition pathways.

\begin{figure}[t!]
\includegraphics[width=0.95\linewidth]{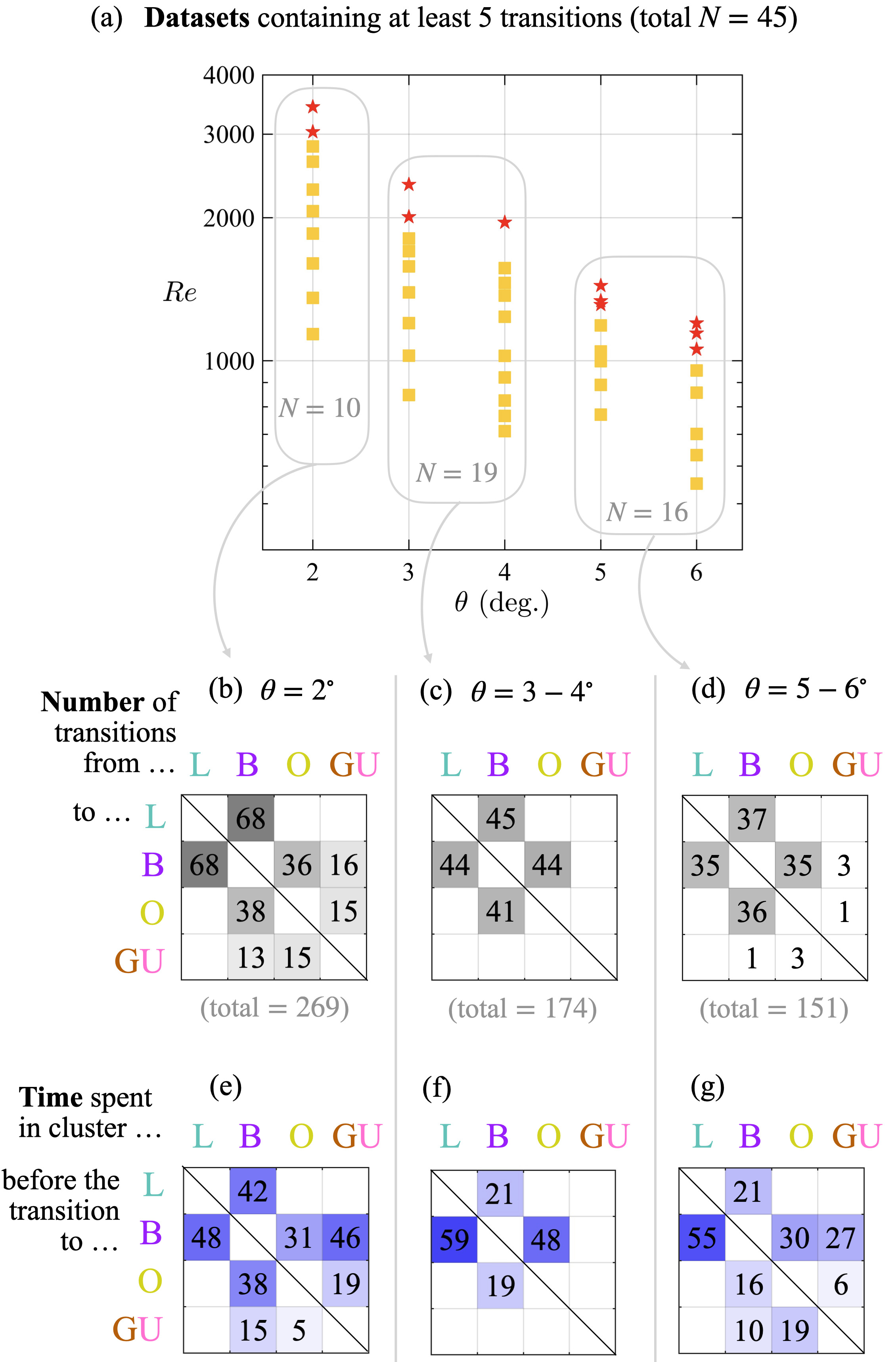}
\caption{Transition and residence matrices computed by the cluster-based network model. (a) The 45 intermittent experiments used for the analysis, exhibiting at least five transitions each. (b-d) Number of transitions $N_{X\rightarrow Y}$ recorded and (e-g) average residence times $T_{X\rightarrow Y}$ for all data with tilt angles $\theta=2^\circ,\, 3-4^\circ, \, 5-6^\circ$. }
\label{fig:matrices}
\end{figure}

\textbf{Cluster-based network model of intermittency.} -- We now study these intermittent cycles systematically in fig.~\ref{fig:matrices} with a cluster-based network model inspired by
\cite{li_cluster-based_2021}. We count, for each experiment, the number of transitions $N_{X\rightarrow Y}$ between clusters X and Y  and the residence times $T_{X\rightarrow Y}$ (in advective time units) spent by the flow in cluster X before transitioning to cluster Y. This  model overcomes the  limitations of traditional Markov chain models \cite{kaiser_cluster-based_2014,foroozan_unsupervised_2021} by providing residence times rather than selecting a fixed arbitrary time step to discretise the time series. We restrict the analysis to the $45$ truly intermittent experiments exhibiting at least five transitions (fig.~\ref{fig:matrices}a), comprising a total of 594 transitions.
Three extra steps were taken for robustness. First, we assigned all unclustered data points (in black in fig.~2a) to their nearest cluster. Second, we merged the neighbouring clusters G and U into a single GU cluster given their similarity. Third, we denoised the data by excluding transitions near cluster boundaries of the form X$\rightarrow$Y$\rightarrow$X that are deemed spurious below a short residence of $T_{Y \rightarrow X}<3.5$, instead treating the data as remaining in cluster X. Figure \ref{fig:matrices}b-d show the $N_{X\rightarrow Y}$ matrices by segregating all the data with tilt angle $\theta=2^\circ$ ($N=10$ experiments),  $\theta=3-4^\circ$ ($N=19$) and  $\theta=5-6^\circ$ ($N=16$). We find that these three matrices are nearly symmetric, showing a similar number of transitions from X$\rightarrow$Y and Y$\rightarrow$X. The most common types of transitions are L$\leftrightarrow$B (50\% of the total 594), followed by  B$\leftrightarrow$O (39\%). The transitions B$\leftrightarrow $GU and O$\leftrightarrow$GU are rarer (5.5\% each) and almost exclusively found at $\theta=2^\circ$, confirming the different routes to turbulence found in fig.~4. We do not observe any direct transitions between the laminarising (L) and turbulent (O and GU) clusters, confirming the role of the braided (B) cluster as the unique gateway to and from intense turbulence.  Figure \ref{fig:matrices}e-g show the corresponding averaged residence matrices $T_{X\rightarrow Y}$, and we pay particular attention to their  asymmetry. We find that residence times on either side of the L$\leftrightarrow$B transitions are nearly equal at 
$\theta=2^\circ$ (at 42 and 48 time units). By contrast, at $\theta=3-6^\circ$, the flow spends much less time in B before L than vice versa ($T_{B \rightarrow L}\ll T_{L \rightarrow B}$), signalling a faster relaminarisation at higher tilt angles for reasons that remain to be understood. We also find that  $T_{B \rightarrow X}\ll T_{X \rightarrow B}$ (with only one exception for $T_{B \rightarrow O}$ at $\theta=2^\circ$, which will be minimised later), meaning that B is a relatively unstable region of phase space, transitioning mostly to either L or O based on fig \ref{fig:matrices}b-d.  We also notice a strong asymmetry in times between turbulent clusters O$\leftrightarrow$GU, as
$T_{O\rightarrow GU}\ll T_{GU \rightarrow O}$ (5 vs 19) at $\theta=2^\circ$, whereas it is the exact opposite at $\theta=5-6^\circ$ (19 vs 6). This is a puzzling but potentially insightful observation which may require a greater sample of transitions to be confirmed as statistically significant.

\begin{figure}[t]
\includegraphics[width=0.999\linewidth]{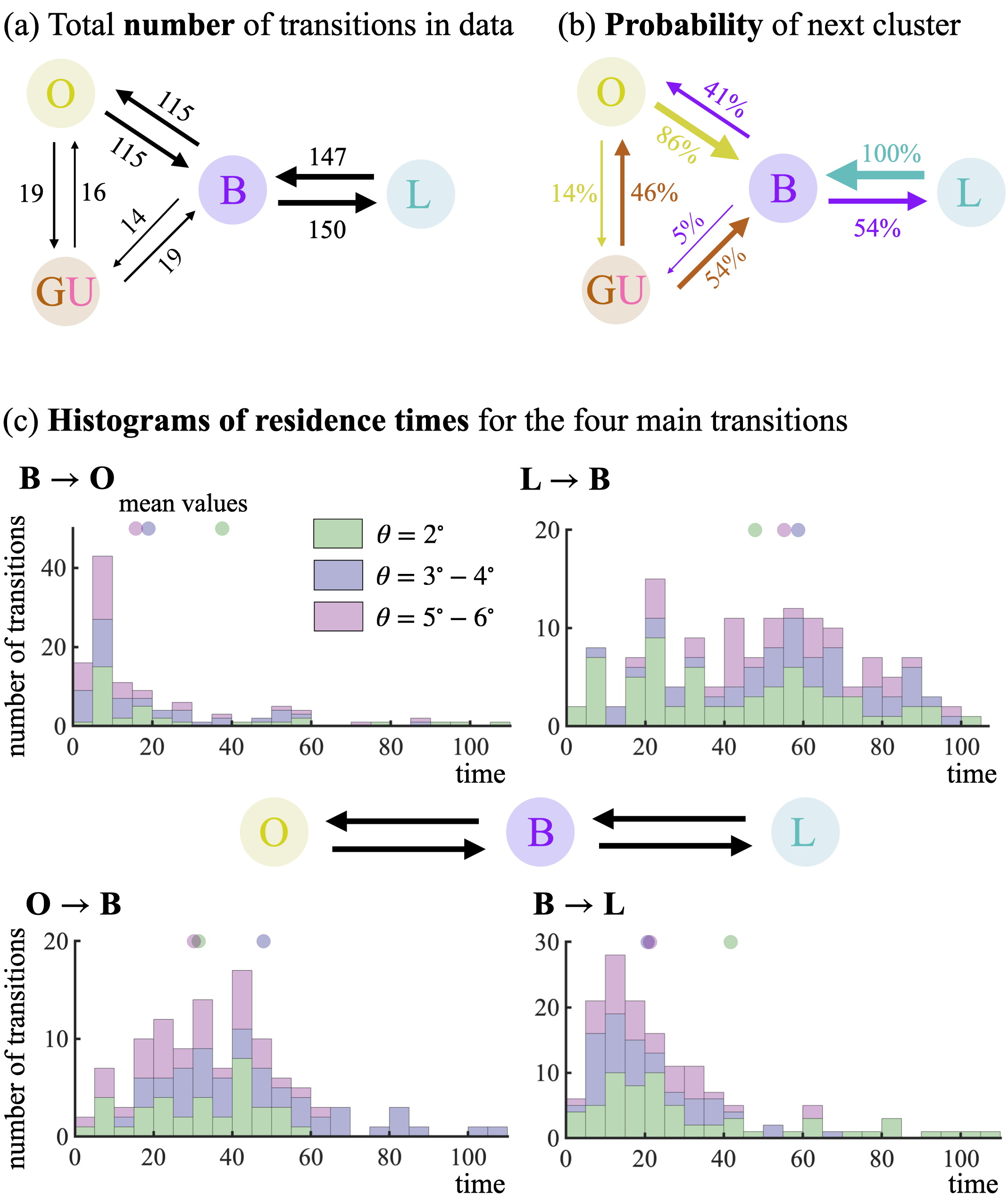}
\caption{Intermittency statistics from the network model. (a) Total number of transitions. (b) Normalised `next-cluster' probabilities (the coloured arrows out of each cluster sum to 100\%). (c) Probability distribution of the residence times for the B$\leftrightarrow$O and B$\leftrightarrow$L transitions, coloured by range of $\theta$. The mean values, given by the small circles on top of each histogram, match the matrices of fig.~5e-g.}
\label{fig:histograms}
\end{figure}

\textbf{Probabilistic description of transitions.} -- Statistics on transitions and residence times are shown in fig.~\ref{fig:histograms} to offer a condensed, low-dimensional representation of SID intermittency. The total number and type of transitions in the data (fig.~\ref{fig:histograms}a) may be normalised to provide, for each cluster, the probability of transition to the next cluster (fig.~\ref{fig:histograms}b). 
We find that laminarising (L) turbulence always leads to braided (B) turbulence, while overturning (O) turbulence most often transitions back to braided (B) turbulence. By contrast, the task of predicting the future state of the flow from a single snapshot in granular/unstructured (GU) or braided (B) turbulence would be more uncertain as their probabilities are more evenly spread across two or three clusters. Figure~\ref{fig:histograms}c shows the probability distributions of residence times of the four most common transitions O$\leftrightarrow$B$\leftrightarrow$L. We find the most spread for laminarising turbulence (L$\rightarrow$B, top right panel), and the least spread for braided turbulence before it overturns (B$\rightarrow$O, top left panel). However, we notice a few outliers with very long $T_{B \rightarrow O}$, especially at $\theta=2^\circ$ (in green) which increase the mean value above those at higher angles, resulting in the apparent exception previously noted in fig.~5.  Outliers are also found in $T_{B \rightarrow L}$ (bottom right) at $\theta=2^\circ$, suggesting that braided turbulence is occasionally surprisingly long-lived  at low tilt angles. These histograms also show that the vast majority of overturning turbulence events last less than 70 advective (shear) time units ($T_{O \rightarrow B}\le70$), and laminarising phases last less than 100 time units ($T_{L \rightarrow B}\le100$). What sets these robust upper bounds in times scales remains unclear, but it may be related to the length of the  duct. As the longitudinal aspect ratio is 40, it takes $\approx 80$ times units for a turbulent patch to be advected at the layer-averaged velocity along the duct. It must be kept in mind that our data provide a localised view of these spatio-temporally intermittent dynamics in a window spanning around a fifth of the duct length.

\textbf{Trends in residence times under increasing turbulence.} -- We finally  turn our attention in fig.~7 to the evolution of residence times as the dynamic range $\textrm{Re}\, \theta$ is increased within the intermittent region of parameter space $\textrm{Re}\, \theta\approx 3000-7000$. We focus on the numerous transitions between braided and overturning turbulence, which show the clearest trend. We plot the mean (circle) and minimum/maximum (error bars) $T_{B \rightarrow O}$ (fig.~7a) and $T_{O \rightarrow B}$ (fig.~7b) between $\theta=3-6^\circ$, with larger circles denoting a larger sample of transitions. As sketched by the light blue trend arrow, we find in fig.~7a that the unstable braided periods $T_{B \rightarrow O}$ first decreases with $\textrm{Re}\, \theta$ in the range $3000-5000$ (see grey inset) to very short and repeatable below $5$  time units before increasing again to generally longer and much more variable values of order $10-50$ at higher $\textrm{Re}\, \theta\approx 5000-7000$. The existence of an optimal, intermediate $\textrm{Re}\, \theta$ minimising the transition time to turbulence is particularly compelling. By contrast, we find in fig.~7b that the overturning turbulence period $T_{O \rightarrow B}$ increases monotonically with $\textrm{Re}\, \theta$. This suggests that the strongly dissipative overturning state becomes more attractive and stable at higher $\textrm{Re}\, \theta$, consistent with $\textrm{Re}\, \theta$ being a proxy for the mean dissipation rate. Remarkably, both panels of fig.~7 demonstrate a  collapse of residence times with $\textrm{Re}\, \theta$ at various values of $\theta$. This product of parameters is therefore not only significant to the first-order energetics and the distribution of types of turbulence in parameter space, but also to more subtle and deeper statistical properties of its temporally intermittent dynamics in phase space.

\begin{figure}[t]
\includegraphics[width=\linewidth]{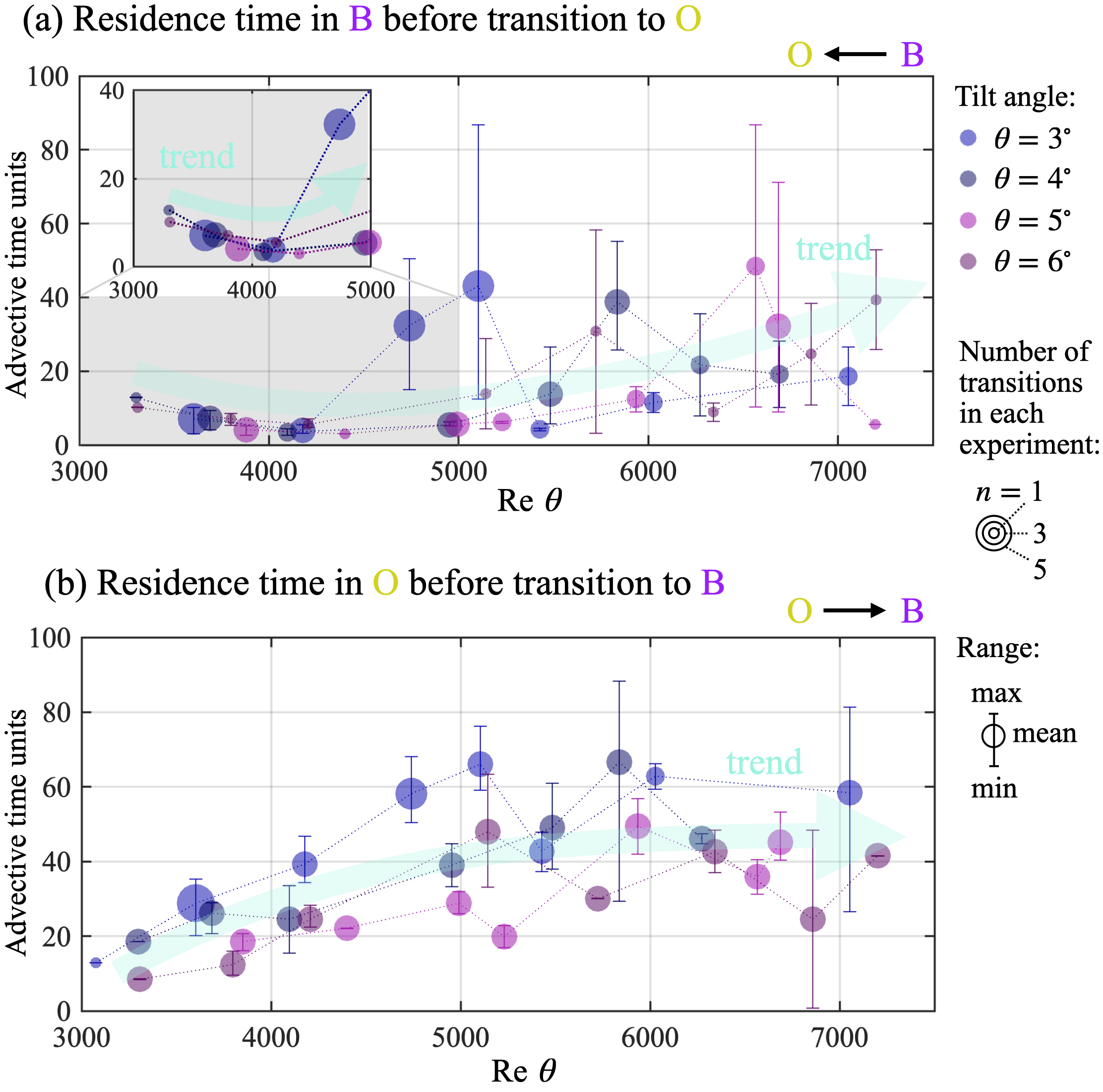}
\caption{Trends in residence times (a) $T_{B \rightarrow O}$ and (b) $T_{O \rightarrow B}$ with increasing turbulent dynamic range (or dissipation) $\textrm{Re}\, \theta$, using the 35 intermittent experiments of fig.~5a between $\theta=3-6^\circ$. Data at $\theta=2^\circ$ are excluded because L$\leftrightarrow$B$\leftrightarrow$O cycles are replaced by L$\leftrightarrow$B$\leftrightarrow$UG cycles at higher $\textrm{Re}$ values, as found in fig.~4. Note the emphasis (larger circles) on experiments having a larger, more statistically significant sample of transitions (maximum 5). }
\label{fig:OB_plots}
\end{figure}

\textbf{Summary and conclusions.} --- We have performed  coherent structure modelling of turbulence in the SID experiment, a wall-bounded sustained stratified shear flow whose  2-D parameter space $(\textrm{Re},\theta)$ yield a rich set of turbulent and intermittent states. Our analysis was made possible by a recent image-processing algorithm that transforms shadowgraph movies into a reduced set of 2-D vectors representing the morphology of density interfaces within each frame. This allowed an unsupervised algorithm to automatically reveal five distinct types of turbulence, interpreted as building blocks for the coherent `skeleton' underpinning SID turbulence, and map their distribution in parameter space. 
The temporal dynamics of turbulence with high  dissipation and dynamic range $\propto \textrm{Re} \, \theta$ are confined within individual clusters, whose type shifts from unstructured to granular to overturning with decreasing $\textrm{Re}$ and increasing $\theta$. Less dissipative, intermittent turbulence at lower $\textrm{Re} \, \theta$ transitions between clusters, typically quasi-periodically. Two fundamentally different routes to turbulence were identified, both of which pass through a common braided turbulence `gateway' but end up in different granular or overturning `attractors'.
We developed a dynamic network model to quantify cluster transition probabilities  and residence times in a dataset of 594 transitions. This low-dimensional  description of SID intermittency revealed the following physical insights. (i) Average residence times greatly vary between clusters and, within each cluster, depend on the next cluster visited. (ii) The braided turbulence gateway is the most unstable region, and relaminarisation is faster at higher $\theta$. (iii) Predicting the next cluster is more uncertain from a single snapshot in braided, granular or unstructured turbulence compared to overturning or laminarising turbulence, but residence times in braided turbulence are the most predictable. (iv) 
Intense turbulence and laminarising phases tend to be shorter than 100 advective (shear) time units, which may be related to advection along the duct. (v) The transition to turbulence is, surprisingly, shortest at intermediate $\textrm{Re} \, \theta$. By contrast, the dissipative overturning turbulence attractor becomes increasingly stable with $\textrm{Re} \, \theta$ until intermittency disappears, with similarities to directed percolation in pipe flow \cite[\S~10.4]{barkley_theoretical_2016}. This provides further evidence that the bulk dissipation proxy $\textrm{Re} \, \theta$ has a deep and subtle influence on the geometry of the SID `skeleton' in phase space that warrants further study.  The success of this reduced-order modelling also suggests that a similar methodology could profitably be applied to gain new dynamical systems insight from other datasets of high-$\textrm{Re}$ turbulence with additional physics, such as rotating, multiphase, or magnetohydrodynamic turbulence.

\acknowledgments

We thank  G. Kong, S. B. Dalziel and the technicians of the GK Batchelor Laboratory for their help with the experiments, and C. P. Caulfield for insightful discussions. The facility and XJ were funded by an ERC Horizon 2020 Grant No. 742480 led by P. F. Linden. AL acknowledges a Leverhulme Early Career Fellowship and a NERC Independent Research Fellowship (NE/W008971/1). YHMC acknowledges  the Summer Research in
Mathematics  internship programme and Magdalene College. MMPC acknowledges  the Natural Sciences and Engineering Research Council of Canada (NSERC) through RGPIN-2024-06184. 
The shadowgraph data  are available at \cite{jiang_research_2023_EPL} and the reduced and clustered data  are available at \cite{lefauve_research_2024_EPL} .


\end{document}